\documentclass[iop,apj]{emulateapj}
\usepackage{graphicx}
\usepackage{epsf}
\usepackage{natbib}
\usepackage{amsmath}
\usepackage{color}
\usepackage{times}
\usepackage{microtype}
\usepackage{float}
\usepackage[T1]{fontenc}
\usepackage{pslatex}
\def\la{\mathrel{\hbox{\rlap{\hbox{\lower4pt\hbox{$\sim$}}}\hbox{$<$}}}}
\def\ga{\mathrel{\hbox{\rlap{\hbox{\lower4pt\hbox{$\sim$}}}\hbox{$>$}}}}

\def\kms{km~s$^{-1}$}

\def\dm15{{$\Delta$}$m_{15}$}

\def\v10{$V_{10}$(Si~II)}

\def\W575{$W(5750)$}
\def\W610{$W(6100)$}
\def\6100{the 6100~\AA\ absorption}

\def\msun{M$_\odot$}

\def\ni{$^{56}$Ni}
\def\co{$^{56}$Co}

\def\gcm3{g~cm$^{-3}$}
\def\cm2g{cm$^{2}$~g$^{-1}$}

\def\CaII7291{[Ca~{\sc II}] $\lambda\lambda$7291,7323}

\def\OI6300{[O~{\sc I}] $\lambda\lambda$6300,6364}

\def\apj{ApJ}
\def\apjl{ApJL}

\def\nat{Nature}
\def\mnras{MNRAS}

\begin{document}

\title{{\bf Circumstellar Interaction Models for the Bolometric Light Curve of SN~2017egm}}

\author{J. Craig Wheeler\altaffilmark{1}, Emmanouil Chatzopoulos\altaffilmark{2}, Jozsef
  Vink{\'o}\altaffilmark{1,}\altaffilmark{3,}\altaffilmark{4}, Richard Tuminello\altaffilmark{2}}
\email{wheel@astro.as.utexas.edu}
\altaffiltext{1}{Department of Astronomy, University of Texas at Austin, Austin, TX, USA}
\altaffiltext{2}{Department of Physics \& Astronomy, Louisiana State University, Baton Rouge, LA, 70803, USA}
\altaffiltext{3}{Konkoly Observatory, Research Center for Astronomy and Earth Sciences of
the Hungarian Academy of Science, Konkoly Thege M. ut 15-17, Budapest, 1121,
Hungary}
\altaffiltext{4}{Department of Optics and Quantum Electronics, University of Szeged,
Dom ter 9, Szeged, 6720, Hungary}

\begin{abstract}

We explore simple semi-analytic fits to the bolometric light curve of Gaia17biu/SN~2017egm,
the most nearby hydrogen-deficient superluminous supernova (SLSN~I) yet discovered. 
SN~2017egm has a quasi-bolometric light curve that is uncharacteristic of other SLSN~I by 
having a nearly linear rise to maximum and decline from peak, with a very sharp transition. 
Magnetar models have difficulty explaining the sharp peak and may tend to be too bright 
20 d after maximum. Light curves powered only by radioactive decay of \ni\ fail on similar 
grounds and because they demand greater nickel mass than ejecta mass. Simple models 
based on circumstellar interaction do have a sharp peak corresponding to the epoch when 
the forward shock breaks out of the optically-thick circumstellar medium or the 
reverse shock reaches the inside of the ejecta. We find that models based on circumstellar 
interaction with a constant-density shell provide an interesting fit to the bolometric light curve 
from 15 d before to 15 d after peak light of SN~2017egm and that both magnetar 
and radioactive decay models fail to fit the sharp peak. Future photometric observations
should easily discriminate basic CSI models from basic magnetar models. The implications
of a CSI model are briefly discussed.

\end{abstract}


\section{Introduction}
\label{intro}

The first identified hydrogen-deficient superluminous supernova (SLSN~I), 
SN~2005ap, was discovered by the Texas Supernovae Search \citep{quimby07}. 
SLSN~I are now recognized as a distinct class \citep{quimby11} that can be 
identified both from their bright light curves and their spectra. Their progenitor 
evolution and the source of their great optical luminosity remain uncertain. Most 
SLSN~I have appeared in dwarf galaxies of high star formation rate and low 
metallicity \citep{quimby11,neill11,stoll11,chen13,lunnan14,perley16}, even 
specifically in extreme emission line galaxies \citep{leloudas15}.

Gaia17biu = SN~2017egm was discovered by the Gaia mission on 2017 May 23.
It was 
subsequently classified
as an SLSN~I \citep{dong17,nicholl17a}. The host galaxy, NGC~3191, is atypical
of SLSN~I hosts, being massive with a mean metallicity near solar. This raises
issues as to whether SLSN~I only form in low metallicity and, if so, what is
the metallicity cutoff above which SLSN~I do not form 
\citep{nicholl17a,bose17,chen17,izzo17}.

The light curve of SN~2017egm also has remarkable properties. Its peak
brightness is on the low end of the distribution of SLSN~I \citep{decia17,lunnan17}
with a maximum of M $\approx$ -21. Even more interesting, perhaps, is the
nature of the light curve. In the approximately 20 d before peak and in the 
subsequent first 20 d of decline, the individual bands are nearly linear in magnitude 
and hence exponential in time \citep{bose17}. The compiled bolometric light curve 
that spans a somewhat shorter time is also nearly linear on the rise and decline. 
Upon closer inspection (Figure \ref{minim}) the rise and decline near peak are both 
concave with positive second derivative. V-band data prior to 20 d before maximum
shows the more familiar negative second derivative. The peak itself is 
unprecedentedly sharp, separating the quasi-linear rise from the quasi-linear 
decline \citep{bose17}. Many models intrinsically fail to give that shape; both 
input from radioactive decay and from a magnetar have negative second derivatives 
on the rise and rounded peaks. \citet{nicholl17a} have successfully fit the rise to 
peak with a magnetar model, but they did not have access to post-peak photometry 
and hence did not attempt to fit the sharp peak nor the subsequent tail. Comparing 
the models of \citet{nicholl17a} to the data of \citet{bose17} shows that the
dipole-driven magnetar models do not reproduce the sharp peak and hint that the 
magnetar models are somewhat too bright by 20 d after maximum. 

Simple quasi-analytic light curve models based on circumstellar interaction (CSI)
in spherical geometry naturally give ``kinks" in the computed light curve
\citep{chatz12,chatz13}. These breaks in the slope of the light curve arise when 
the forward shock reaches the point where the diffusion time becomes less than 
the dynamical time of the forward shock and when the reverse shock reaches the 
inside of ejecta. 
Here, we explore that possibility.

In \S\ref{models} we outline the models as presented by \citet{chatz12,chatz13}.
Section \ref{results} gives our results and \S\ref{discuss} presents our results
and conclusions.

\section{Models}
\label{models}

We have employed two codes to search for fits to light curve data that
minimize the $\chi{^2}$ per degree of freedom for a given model. 
The models can be hybrid, employing the physics of radioactive decay, 
of power from a magnetar, and from CSI. One code is {\tt MINIM}, as 
described in \citet{chatz13}. The other is a new variation, {\tt TigerFit}, a 
{\tt Python} code developed by E.C. that utilizes the {\tt Numpy} and 
{\tt SciPy} packages and, in particular, the {\tt SciPy.optimize.curve\_fit} 
method to search through a grid of parameters and determine the best--fit 
model via the minimization of the $\chi^{2}$--statistic. {\tt TigerFit} accepts 
as input the rest--frame pseudo--bolometric light curve of a supernova or 
other transient event and can be asked to fit a variety of semi--analytic light 
curve models for different power inputs based on the method of \citet{arnett82}. 
Some of these models are outlined in \citet{chatz12}. 
{\tt TigerFit} 
is open--source software and can be obtained from a GitHub repository 
online\footnote[1]{https://github.com/manolis07gr/TigerFit}.

{\tt MINIM} and {\tt TigerFit} were designed to give quick constraints on
light curves, especially those with unexpected properties, by searching 
through the parameters of a variety models of input physics. SN~2017egm
provides an excellent example of the application for which these codes
were designed. 

An important limiting assumption of our CSI models is that we assume that 
the effects of the forward and reverse shock heating are both centrally located.
Forward shock heating terminates when the shock breaks 
out of the CSM and reverse shock heating when the whole ejecta mass has 
been swept through \citep{chatz12}. The assumption of a centrally-located power 
source for the case of the forward and the reverse shocks, although convenient, 
is not generally true and thus increases the uncertainties and limitations of this 
approximate model. 
As it turns out, the models presented in \S\ref{results} depend 
mostly on the reverse shock and do not depend sensitively on the behavior of 
the forward shock. 

When energy input declines monotonically with time (for radioactive decay, magnetar,
or a CSI wind model), the light curve on the rise has a monotonically decreasing slope, 
reminiscent of most observed supernovae. When the energy input rises monotonically 
with time, the light curve on the rise has a monotonically increasing slope. This latter
behavior can arise when the reverse shock propagates into a steeply increasing 
density profile \citep{chatz12}. After shock input ceases in the CSI models, the decline 
in luminosity is predicted to be exponential, reflecting diffusion from the expanding 
matter. 
The radioactive decay and magnetar 
models predict their own unique declines. The shape of the rising and declining light 
curve is thus a potentially strong constraint on the models.

In the radioactive decay and magnetar models the parameter $R_p$ 
represents the radius of the progenitor star. In the CSI models, this parameter
serves as the inner radius of the CS material, the inner edge of a shell or the
inner boundary of the wind. Specifically, in the CS shell model, the CSM extends
from $R_p$ to the outer radius of the shell, $R_{CSM}$, and $R_p$ does not
measure the actual radius of the progenitor which is not constrained in this class
of models. At $t=0$, the supernova ejecta are assumed to be in homologous expansion 
and first contacting the CSM shell at $R_p$, implicitly assuming that the ejecta catch
up with the shell shortly after explosion. For instance, if the outermost ejecta expand
with $v \sim 50,000$ \kms it takes $\sim$ 5 hours for the ejecta to expand to $10^{14}$ cm, 
even if the progenitor were very compact at the moment of explosion.

We have used the data for the bolometric light curve of SN~2017egm
from \citet{bose17} (their Figure 7). We omitted the first two 
data points that seem to form a short plateau. If these points are real, they 
cannot be modeled with our codes.

\section{Results}
\label{results}

We explored a variety of hybrid models. 
Within the range of parameters of CSI, we can successfully fit models
based on pure CSI. The shape of the rising part of the curve, whether
it is concave or convex depending on the sign of the second derivative
depends on choices of the power law index, $n$, of the outer supernova
ejecta density profile and the power law index, $s$, that characterizes
the density profile of the CSM. In the current set of models we have
only explored the CSM density profiles corresponding to $s = 0$,
constant density, and $s = 2$ corresponding to a steady-state wind with
a profile $\rho \propto r^{-2}$. The models are formally scaled with a wind 
velocity, $v_w$, of $10$ \kms. All models assumed $\kappa = 0.2$ 
cm$^2$ g$^{-1}$ corresponding roughly to a hydrogen-deficient plasma. 
The slope of the outer ejecta density profile corresponded to $n=11$
or $n=12$. The models also contain a small, inner region of constant density. 
For the RAD and MAG models, we assumed an outer expansion velocity of 
20,000 \kms\ \citep{nicholl17a,bose17}.
 
\begin{figure}
\begin{center}
\includegraphics[angle=0,width=3.5 in]{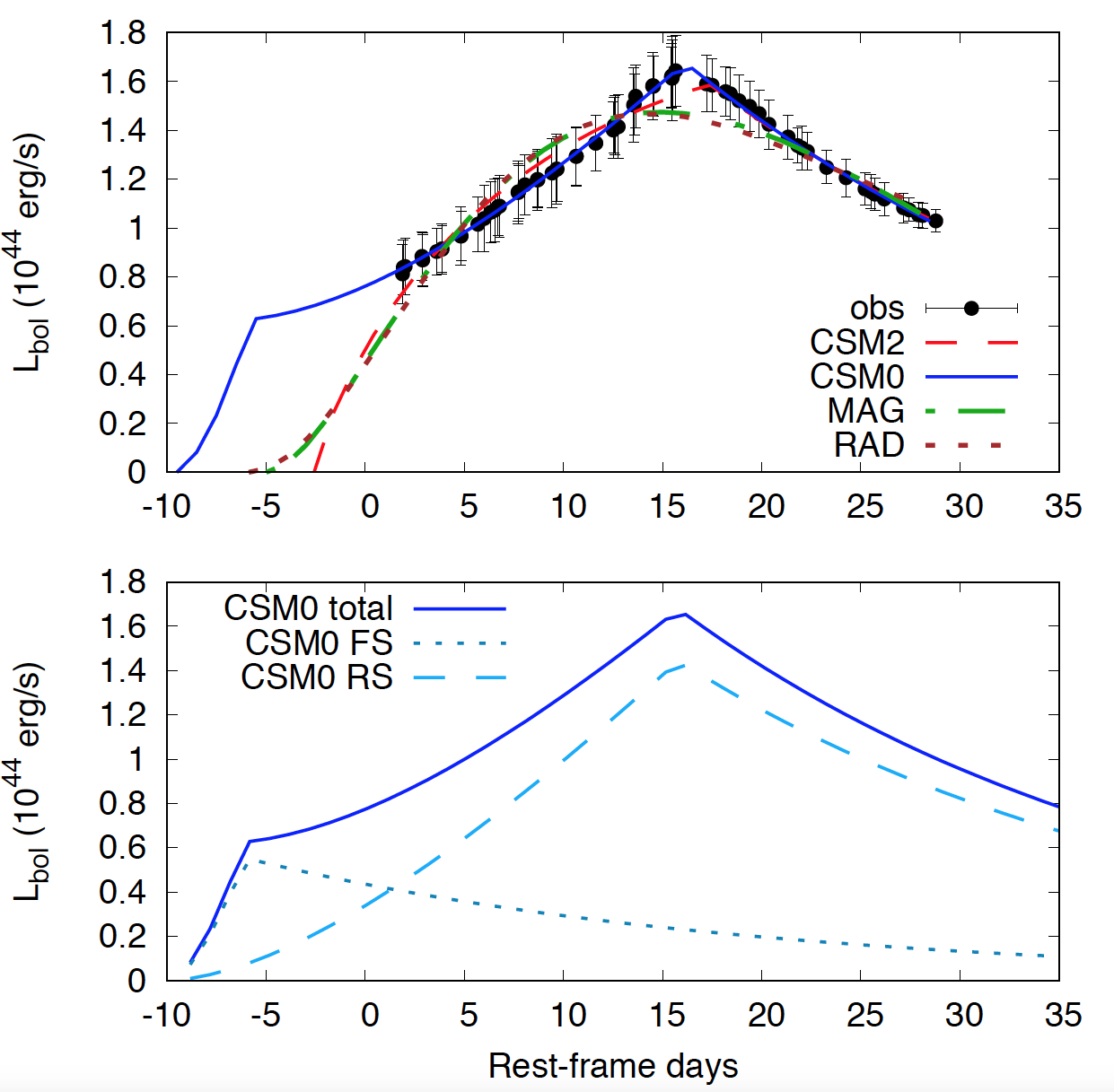}
\caption{{\sl Upper Panel} Bolometric light curve of SN~2017egm from \citet{bose17}, 
their figure 7 (black points). The zero of the time axis is arbitrary and the first two 
data points were omitted (see text). Models computed with {\tt MINIM} are shown with 
input from radioactive decay of  \ni\ and \co\ (RAD; brown dash) and from magnetar input 
(MAG; green long/short dash). Also shown are models based on CSI with two density profiles 
corresponding to a steady state wind $s = 2$ (CSM2; red dash) and to a constant 
density $s = 0$ with an outer cutoff (CSM0; blue). Model parameters are given in Table
\ref{T1}. The constant-density CSI model gives the best fit overall. {\sl Lower Panel}
Decomposition of the {\tt MINIM} best-fit constant density CSI model from Figure the
upper panel showing the effect of the forward shock (green dash) the reverse shock 
(blue dash) and total luminosity (red). The zero point of the time axis is arbitrary.
See the online version for color.}
\label{minim} 
\end{center}
\end{figure}

The upper panel of Figure \ref{minim} shows the light curve fits using {\tt MINIM} 
for pure CSI models with $s=0$ (CSM0) and $s=2$ (CSM2) and models employing 
only radioactive decay (RAD) or magnetar (MAG) input (note the linear scale 
in the figure). The zero point of the time axis in Figure \ref{minim} is arbitrarily 
set to about the time of the first data. The actual explosion and peak times 
vary with the model. The first two points from \citet{bose17} have been dropped.  
The parameters of the models are given in Table \ref{T1}. 

The CSM0 model corresponding to $s = 0$ provides a remarkable, and surprisingly, 
good fit, with a slightly curving rise of a factor of two in flux starting about 15 days 
prior to the peak, a sharp peak, and a decline that reasonably captures the first 
$\sim$ 15 days of decline. Note that both the model and the data formally show 
a slight decrease in slope on the decline. This model required an ejecta of 30 \msun\ 
colliding with a CSM of 0.8 \msun\ with an energy of $\sim 6\times10^{51}$ erg. 
The inner radius of the CSM shell was $9\times10^{13}$ cm. The CSM had an outer 
shell radius of $3.5\times10^{14}$ cm and a density of $8.9\times10^{-12}$ \gcm3.

The lower panel of Figure \ref{minim} shows the decomposition of the best-fitting CSM0 
model of {\tt MINIM}. The explosion occurs about 25 days prior to the peak. The forward
shock reaches the outer edge of the CSM shell about 3 days after the explosion.
After that, the contribution of the forward shock declines exponentially. The reverse
shock produces a steadily increasing energy input as it propagates up the steep
$n = 12$ density profile. Note that this increasing input produces the concave
component that is critical to accounting for the increasing slope of the pre-maximum
light curve. The break to smaller slope at about 2 days before maximum is when
the reverse shock encounters the small inner component of the ejecta with 
assumed constant density. The reverse shock reaches the interior of the ejecta
at maximum. The light curve subsequently declines exponentially, dominated
by diffusion from the matter heated by the reverse shock with a small continuing
contribution from the shell matter heated by the forward shock. 

The CSM0 model gave a fit to the outer velocity of the ejecta of $\sim 50,000$ \kms. 
This is formally in contrast to the observed early photospheric velocity of $\sim 20,000$ 
\kms\ \citep{nicholl17a,bose17}. In the CSM0 model, the first spectrum was obtained 
$\sim$12 days after explosion, $\sim$ 9 days after the breakout of the forward shock.
The shell would thus have been fully shocked and homologously expanding at the 
epoch of the first spectrum. If half the energy of the explosion, $\sim 6$ foe, were 
delivered as kinetic energy to the shell, the expansion velocity of the shell would have 
been $\sim 19,000$ \kms, close to the observed value. At this early phase, the 
photosphere should still be in the shocked shell, so the CSM0 model may be roughly 
in agreement with the observed velocity. 

The model with $s=2$ gave a somewhat poorer fit, but also had a sharp peak. 
While formally comporting with the constraints of the error bars, the RAD and MAG 
models are less satisfactory with a larger $\chi^2/\text{d.o.f}$ and clearly cannot reproduce 
the sharp peak demanded by SN~2017egm. Our estimates for the properties of magnetar
models are consistent with those of \citet{nicholl17a}. 

\begin{figure*}
\begin{center}
\includegraphics[angle=0,width=7in]{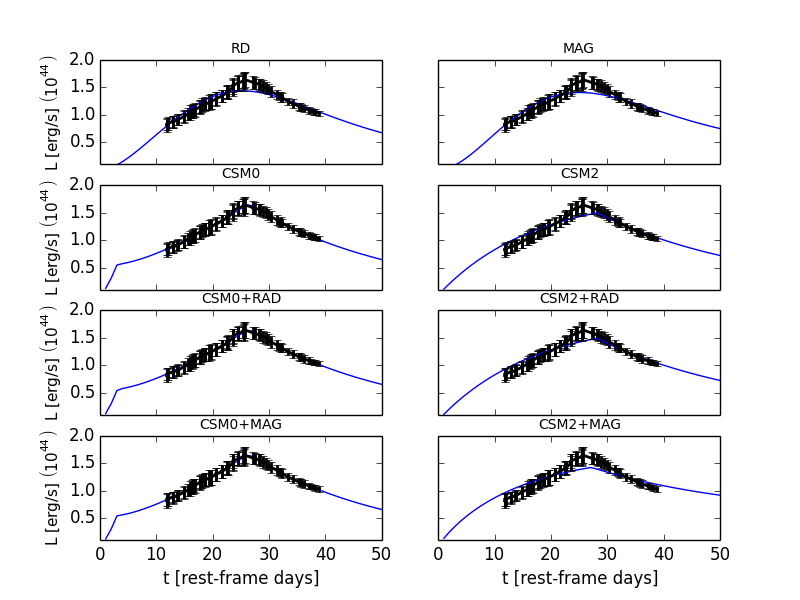}
\caption{Bolometric light curve of SN~2017egm from \citet{bose17} as in 
Figure \ref{minim} (black points). Fits of the \ni\ decay (RAD), magnetar (MAG), 
CSM models with constant density (CSM0) and wind-like profiles (CSM2)
and hybrid models with combined inputs are given by the blue curves. See
text for discussion. See the online version for color.}
\label{tigerfit} 
\end{center}
\end{figure*}

Figure \ref{tigerfit} shows a variety of light curve fits using {\tt TigerFit}
(again excluding the first two data points). The parameters corresponding to 
the results given in Figure \ref{tigerfit} are given in Table \ref{T1}. These 
models formally assume a region of constant density ($\delta = 0$) in the 
inner region of the ejecta. Models based purely on radioactive 
decay (upper left) or on a magnetar (upper right) are again clearly inadequate 
to capture the sharp peak of SN~2017egm. Models with pure CSI (next 
two panels) give sharp peaks and decent fits. The model with constant 
density CSM ($s=0$; left panel) gives an especially good fit. For this model, 
the break when the shock reaches optically-thin regions occurs prior to the 
earliest data. With appropriate choice of parameters, the model peak representing 
the break in slope when the reverse shock reaches the interior of the ejecta fits the 
data very well. This model requires an inner-shell radius of 
$\sim 6\times10^{13}$ cm for a hydrogen-poor opacity of 0.2 cm$^2$ g$^{-1}$
with a mass of $\sim 30$ \msun\ exploding with an energy of $\sim 5\times10^{51}$
erg colliding with a CSM of $\sim 0.6$ \msun. The CSM shell has an outer 
radius of $2.1\times10^{14}$ cm corresponding to a density of 
$3.1\times10^{-11}$ \gcm3. The parameters of this fit are very similar to those 
derived from the pure CSI model based on {\tt MINIM} given in Table \ref{T1}. 
The model with a wind-like CSM profile (right panel) again gives a somewhat poorer fit. 

Models with constant density CSM and a modicum of decay or magnetar input 
(lower two left-hand panels) also give good fits to the observed light curve at
the expense of greater parameter degeneracy. These models formally
allowed about 1 \msun\ of \ni\ and a field of $\sim 10^{14}$ G, respectively. 
In these cases, the break to less steep slope when the shock reaches 
optically-thin depths in the CSM and the break when the reverse shock reaches 
the inner limits of the ejecta both occur near maximum light in a manner 
that still approximately captures the sharp peak observed in SN~2017egm. The 
hybrid models with wind-like CSM profiles (lower two right-hand panels) provide 
less good fits. In addition to the poor fit, the model powered solely by radioactive 
decay (RAD) is unphysical because it requires substantially more \ni\ than the ejecta 
mass. The pure magnetar model (MAG) requires a rather small ejecta mass. The 
model with $s=2$ CSM and magnetar input rises after $\sim 40$ d because the 
effects of the CSM input die out while the magnetar input continues for this particular 
model that otherwise showed the lowest $\chi^{2}/\text{d.o.f.}$ value in {\tt TigerFit} 
for this class of models. 

\section{Discussion and Conclusions}
\label{discuss}

The nearby SLSN~I SN~2017egm displays a rather special quasi-bolometric light 
curve for the 15 d before and after peak with a sharp peak separating the linear rise
from the linear decline. Both radioactive decay models and magnetar models give
rounded light curves with smooth peaks. In contrast, models based on CSI
can, in principle, yield roughly linear rise and decline joined at a sharp peak.
We note that other treatments can give smoother transitions between epochs 
of CSI models \citep{moriya13}. In our models, the eruption of the forward shock from the 
CS shell causes an early small break in the slope of the light curve. In our models, 
the abrupt change in slope at peak light corresponds to the epoch when the reverse 
shock reaches the innermost ejecta.  The subsequent lingering exponential
decline comes from the continued diffusive release of energy from deeper layers. 
We have presented several models showing that CSI can account for the light curve 
shape of SN~2017egm around maximum light, including models in which the CSI is 
abetted by a modest input from radioactive decay or a magnetar. 

\begin{figure*}
\begin{center}
\includegraphics[angle=0,width=3.5in]{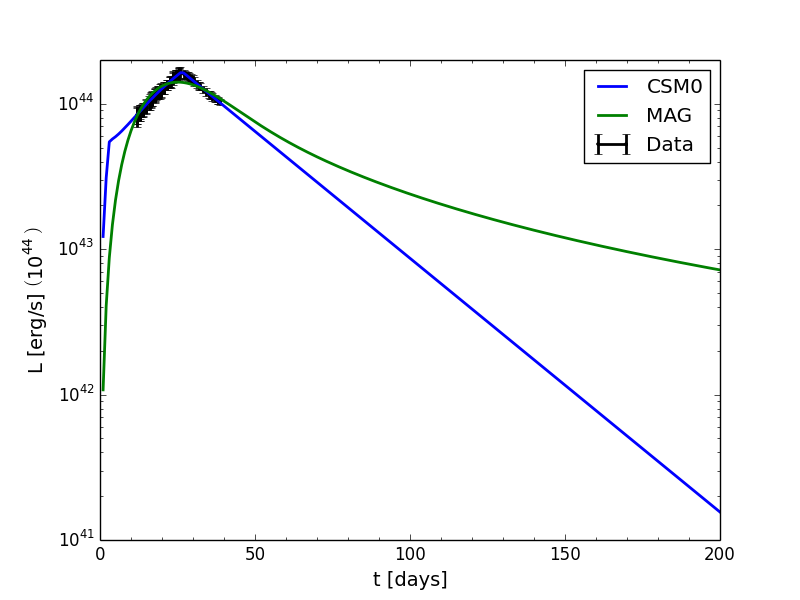}
\caption{Bolometric light curve of SN~2017egm from \citet{bose17} as in 
Figure \ref{minim} (black points) with best-fit CSI and magnetar models 
from Figure \ref{tigerfit} extrapolated to longer time periods. The diffusive
exponential decline of the CSI model (blue) is predicted to fall off much more
steeply than the power law of the simple magnetar model (red).  See the
online version for color.}
\label{contrast} 
\end{center}
\end{figure*}

The data currently available to us mitigate against pure radioactive decay or
pure magnetar models, but the data only span about 30 days around peak. A 
critical test will be the subsequent behavior of the light curve. The bolometric
light curve of a radioactive-decay model follows certain systematics driven
by the physics of the weak interactions. At late times, the bolometric light
curves of basic dipole-driven magnetar models are predicted to decay like
$t^{-2}$. Both the decay models and the magnetar models can be altered
by leakage of gamma-rays, positrons, or magnetar input at the expense of
added fitting parameters. CSM models are intrinsically bedeviled by a profusion
of fitting parameters, but they do have characteristics that allow sharp breaks
in the light curve. It is possible that SN~2017egm is displaying some of
those characteristics. Again, more extensive data on the decline will provide
important constraints on CSI models.

Figure \ref{contrast} shows the extrapolation of our best-fit {\tt TigerFit} CSM0 
and MAG models to 200 days. At face value, photometric monitoring should
easily discriminate these two basic models, with CSM0 decaying 
exponentially, and MAG decaying as a power-law, $t^{-2}$. In practice,
variations in both models might decrease the contrast. Our simple MAG 
model ignores the dynamical effects of a wind-blown shell \citep{kasen}
or leakage effects \citep{nicholl17b}. As the luminosity declines, some 
input from \co\ decay might contaminate either model. Nevertheless, the
contrast expected between the two classes of models is very large. 

The data on SN~2017egm and our models suggest that CSM is a major
factor. The issue of whether CSM could be active in SLSN~I despite the
lack of evidence for narrow nebular lines was raised in \citet{chatz12}
and discussed in some more detail in \citet{chatz13}. This remains a
controversial issue, to which SN~2017egm may bring some clarity. 
The slowly-declining SLSN~I are often characterized by irregularities
in the light curve that are most straightforwardly attributed to CSI 
\citep{decia17}. If a class of SLSN or a single example in the case of 
SN~2017egm argue for CSI despite the lack of narrow lines, then the 
implication of the lack of narrow lines must be reconsidered in all SLSN~I. 

\acknowledgments
We are grateful to Subash Bose and Subo Dong for sharing their data
and to Wenbin Lu for fruitful discussions.
This work was supported partly by the project ``Transient Astrophysical
Objects'' GINOP-2-3-2-15-2016-00033 (P.I. Vink{\'o}) of the National Research,
Development and Innovation Office (NKFIH), Hungary, funded by the European
Union and in part by the startup funds of Dr. E. Chatzopoulos at Louisiana State 
University. JCW was supported in part by the Samuel T. and Fern
Yanagisawa Regents Professorship.


\setcounter{table}{0}
\begin{deluxetable*}{lcccccccc}
\tablewidth{0pt}
\tablecaption{Fitting and derived parameters for the light curve models of SN~2017egm}
\tablehead{
\colhead{Parameter} &
\colhead{RAD} &
\colhead{MAG} &
\colhead{CSM0} &
\colhead{CSM2} &
\colhead{CSM0+RAD} &
\colhead{CSM2+RAD} &
\colhead{CSM0+MAG} &
\colhead{CSM2+MAG} 
\\}
\startdata

&&&{\tt MINIM} (Figure \ref{minim}) &&&&&\\

$M_{\rm Ni}$~($M_{\odot}$)                   & 11.2 (0.5) & -- & -- & -- &&&&\\
$M_{\rm ej}$~($M_{\odot}$)                  & 4.61 (0.28) & 4.33 (0.22) & 29.7 (1.7) & 50.2 (6.3) &&&&\\
$E_{\rm SN}$~($10^{51}$~erg)               & 11.1 (0.7) & 10.4 (0.5) & 5.7 (0.3) & 6.6 (0.8) &&&&\\
$P_{0,\rm mag}$~(ms)                             & -- & 3.91 (0.08) & -- & -- &&&&\\
$B_{\rm mag}$~($10^{14}$~G)              & -- & 3.20 (0.07) & -- & -- &&&&\\
$R_{\rm p}$~($10^{13}$~cm)                & -- & -- & 8.9 (0.6) & 9.6 (3.9) &&&&\\
$\dot{M}$~($M_{\odot}$~yr$^{-1}$)     & -- & -- & 0.014 (0.001) & 0.076 (0.003) &&&&
\vspace{0.1 cm}
\\
\hline
\vspace{0.1 cm}

&&&{\tt TigerFit} (Figure \ref{tigerfit}) &&&&&\\

$M_{\rm Ni}$~($M_{\odot}$)                  & 13.5  & -- & --  & --  & 1.0  & 0.7  & --  & --  \\
$M_{\rm ej}$~($M_{\odot}$)                  & 4.0  & 3.4 & 30.0      & 63.7  &  30.0 & 63.7      & 30.5 & 63.9   \\
$E_{\rm SN}$~($10^{51}$~erg)              & 9.6  & 8.2 & 5.0      & 6.0    &  5.0 & 6.0     & 4.9 & 4.2  \\
$P_{0,\rm mag}$~(ms)                             & --  & 4.0  & --   & --   & --  & --  & 6.1  & 3.0  \\
$B_{\rm mag}$~($10^{14}$~G)               & --  & 0.6 & --   & --  & --  & --  & 0.8  &  0.6 \\
$R_{\rm p}$~($10^{13}$~cm)                & --  &  -- & 6.0  & 1.0  &  6.0 &  1.0     & 5.0 & 1.2  \\
$\dot{M}$~($M_{\odot}$~yr$^{-1}$)      & --  & --  & 0.11  & 0.8  &  0.10 & 0.77  & 0.08  &  0.8  \\
$\delta$                                             & -- & --  &  0     &  0    &  0  & 0 &  0     &  0  \\
$n$                                                     & -- & --  & 11    & 12   & 11  & 12       & 11    &  12  \\
$s$                                                     & -- &  --&   0     &  2    &   0 &  2 &  0       & 2   \\
$\chi^{2}/\text{d.o.f.}$                        & 0.61& 0.87 & 0.02  & 0.63 & 0.03 & 0.65 & 0.03 & 1.38  \\
\enddata
\tablecomments{{\it RAD:} radioactive decay diffusion model, {\it MAG:} 
magnetar spin--down model, {\it CSM0:} circumstellar
interaction model with $s =$~0, {\it CSM2:} circumstellar interaction 
model with $s =$~2, {\it CSM0+RAD:} hybrid circumstellar
interaction ($s =$~0) and radioactive decay model, {\it CSM2+RAD:} 
hybrid circumstellar interaction ($s =$~2)
and radioactive decay model, {\it CSM0+MAG:} hybrid circumstellar 
interaction ($s =$~0) and magnetar spin--down model,
{\it CSM2+MAG:} hybrid circumstellar interaction ($s =$~2) and magnetar 
spin--down model.
\label{T1}}
\end{deluxetable*}

\end{document}